\documentclass[3p,times,procedia]{elsarticle}
\usepackage{nupha_ecrc}

\volume{00}

\firstpage{1}

\journalname{Nuclear Physics A}

\runauth{}

\jid{nupha}

\jnltitlelogo{Nuclear Physics A}

\usepackage{amssymb}
\usepackage[usenames, dvipsnames]{color}

\biboptions{sort&compress}

\usepackage[figuresright]{rotating}
\usepackage{xspace}

\usepackage{times}
\usepackage{graphicx}
\usepackage{amssymb}
\usepackage{caption}
\usepackage{subcaption}
\usepackage{slashed}
\usepackage{verbatim}
\usepackage[utf8]{inputenc}
\usepackage{breqn}

\def\lres{L_{\rm res}}
\def\aSC{{\kappa_{\rm sc}}}
\def\be{\begin{eqnarray}}
\def\ee{\end{eqnarray}}
\newcommand{\pythia}{{\sc Pythia}\xspace}
\def\x{{\bf x}}
\def\lres{L_{\rm res}}
\def\rres{R_{\rm res}}
\def\jraa{R_{\rm \tiny{AA}}^{\rm jet}}
\def\hraa{R_{\rm \tiny{AA}}^{\rm had}}

\begin{document}

\begin{frontmatter}



\dochead{XXVIIth International Conference on Ultrarelativistic Nucleus-Nucleus Collisions\\ (Quark Matter 2018)}

\title{A simultaneous understanding of jet and hadron suppression}


\author[a]{J. Casalderrey-Solana}
\author[b]{Z. Hulcher}
\author[c,d]{G. Milhano}
\author[e]{D. Pablos}
\author[f]{K. Rajagopal}

\address[a]{Departament de F\'\i sica Qu\`antica i Astrof\'\i sica \&  ICC,
Universitat de Barcelona, Mart\'{\i}  i Franqu\`es 1, 08028 Barcelona, Spain}
\address[b]{Cavendish Laboratory, University of Cambridge, JJ Thomson Ave, Cambridge CB3 0HE, UK}
\address[c]{LIP, Av. Prof. Gama Pinto, 2, P-1649-003 Lisboa , Portugal}
\address[d]{Instituto Superior T\'ecnico (IST), Universidade de Lisboa, Av. Rovisco Pais 1, 1049-001, Lisbon, Portugal}
\address[e]{Department of Physics, McGill University, 3600 University Street, Montr\'eal, QC, H3A 2T8, Canada}
\address[f]{Center for Theoretical Physics, Massachusetts Institute of Technology, Cambridge, MA 02139 USA
\vspace{-0.2in}}

\begin{abstract}
In the context of the hybrid strong/weak coupling model for jet quenching, we perform a global fit to hadron and jet data in the most central bins both at RHIC and LHC. The qualitative and quantitative success of the analysis is attributed to the fact that the model correctly captures the fact that wider jets lose, on average, more energy than the narrower ones, to which high energy hadrons belong. We show how one can understand the relative jet and hadron suppression by analyzing the jet fragmentation functions, and also discuss the role of plasma finite resolution effects.
\end{abstract}

\begin{keyword}
jet quenching \sep holography \sep jet fragmentation functions


\end{keyword}

\end{frontmatter}


\section{Introduction}

Heavy-ion collisions at RHIC and LHC have revealed much about the dynamics and properties of the quark-gluon plasma (QGP). Most striking, these experiments have demonstrated that at the experimentally accessible range of temperatures, the QGP exhibits strong collective phenomena, and that its rapid expansion and cooling are very well described by relativistic viscous hydrodynamics. The smallness of the shear viscosity over entropy density ratio, extracted via comparisons between hydrodynamic simulations and experimental data, 
suggests that QGP behaves as a strongly coupled liquid. This discovery makes it imperative to probe the substructure of QGP to understand how a strongly coupled liquid emerges from quarks and gluons that are weakly coupled at short distance scales.


In order to probe the substructure of QGP, we study the modification of high-$p_T$ jets, whose partons interact strongly with the medium, deposit some of their energy into the plasma, and leave a wake, which adds its own spectrum of particles to the final reconstructed jet. The increased precision and unprecedented transverse momentum reach of experimental data for both hadron \cite{hadronraa_cms,hadronraa_atlas} and jet \cite{jetraa_cms,jetraa_atlas} suppression allows a comparison to theoretical models across a vast kinematical range. With this kind of phenomenological analysis, we aim to obtain a better understanding of the different high-$p_T$ behavior seen for hadron and jet suppression, as well as to establish a connection between these two observables with a third one that has also been measured, the modification of jet fragmentation functions, in order to help draw a consistent and more complete picture of the physics of jet quenching.



\section{The hybrid strong/weak coupling model}

Based on the observation that jet quenching is a multi-scale process, with both perturbative and non-perturbative aspects present, we resort to a phenomenological hybrid approach that combines weak and strong coupling physics at their corresponding regimes of applicability. The physics of hard parton production and subsequent virtuality evolution can be analyzed with weakly coupled QCD. Complementarily, insights into the physics of a strongly coupled QCD plasma can be obtained from the gauge/gravity duality, which conjectures that a large class of plasmas from non-Abelian gauge theories in $N$ dimensions are dual to black brane spacetimes in at least $N+1$ dimensions. In order to combine the physics of both these regimes into one description, the authors in \cite{Casalderrey-Solana:2014bpa,Casalderrey-Solana:2015vaa} introduced the hybrid strong/weak coupling model, which combines the dominant aspects from both energy scales to describe processes relevant to high energy jet propagation through strongly coupled QGP. 

In this model, parton showers are generated and evolved perturbatively with the Monte Carlo event generator \pythia, and each parton's lifetime, the length of time from its creation to its own splitting in the shower, is estimated by a formation time argument as $\tau= 2E/Q^2$, with $E$ the energy of the parton and $Q$ its virtuality. In between splittings, the energy each particle deposits in the medium follows an energy loss formula derived in \cite{Chesler:2014jva,Chesler:2015nqz} from the dual description of an energetic quark propagating through ${\mathcal N}=4$ supersymmetric Yang-Mills (SYM) theory with a large number of colors and infinite coupling. In the gravity side, the geodesics that constitute the string associated to the quark precipitate below the black hole horizon as the string endpoint falls in the holographic direction, leading to energy loss, or more accurately, leading to a degradation of high energy modes into hydrodynamic excitations, at the following rate:

\begin{equation}\label{CR_rate}
\left. \frac{d E_{\rm parton}}{dx}\right|_{\rm strongly~coupled}= - \frac{4}{\pi} E_{\rm in} \frac{x^2}{x_{\rm therm}^2} \frac{1}{\sqrt{x_{\rm therm}^2-x^2}} \quad , ~~~~~~~~~  x_{\rm therm}= \frac{1}{2\aSC}\frac{E_{\rm in}^{1/3}}{T^{4/3}}
\end{equation}
where $E_{\rm in}$ is the parton's initial energy, and $x_{\rm therm}$ is the string thermalization distance or stopping distance. $\aSC$ is a parameter that depends on the details of the particular gauge theory, which we take as a free parameter to be fixed by comparing the model predictions against hadron and jet $R_{\rm \tiny{AA}}$ experimental data.

In order to fulfill energy-momentum conservation, the hybrid model implements a linearized treatment of the wake generated in the QGP due to the passage of the jet, where the deposited energy fully thermalizes into the hydrodynamic background and results in a linearized perturbation to the hadron spectra after hadronization through the Cooper-Frye prescription~\cite{Casalderrey-Solana:2016jvj}.



The total amount of energy loss of an angularly extended object, such as a jet, has to depend on the number of energy loss sources it contains. Given the inability of the plasma to resolve with arbitrary precision the internal structure of a parton shower, one expects that jet suppression will indeed depend on this so-called finite resolution effect (usually referred to as coherence in pQCD \cite{konrad}). Within our model, by the introduction of a screening length, $\lres \equiv \rres / \pi T$, where $\rres$ is a fixed dimensionless parameter which we argue in \cite{Hulcher:2017cpt} to be $\mathcal{O}(1)$, the effective space-time picture of the shower is modified, such that unresolved dipoles lose energy as one effective emitter. This has the effect of concealing lower energy particles within larger energy effective emitters, reducing the overall jet energy loss.

\section{A fit to central hadron and jet data}

\begin{figure}[t]

	\centering 
	
	\vspace{-0.1in}
	\hspace*{\fill}
	\includegraphics[scale=0.85]{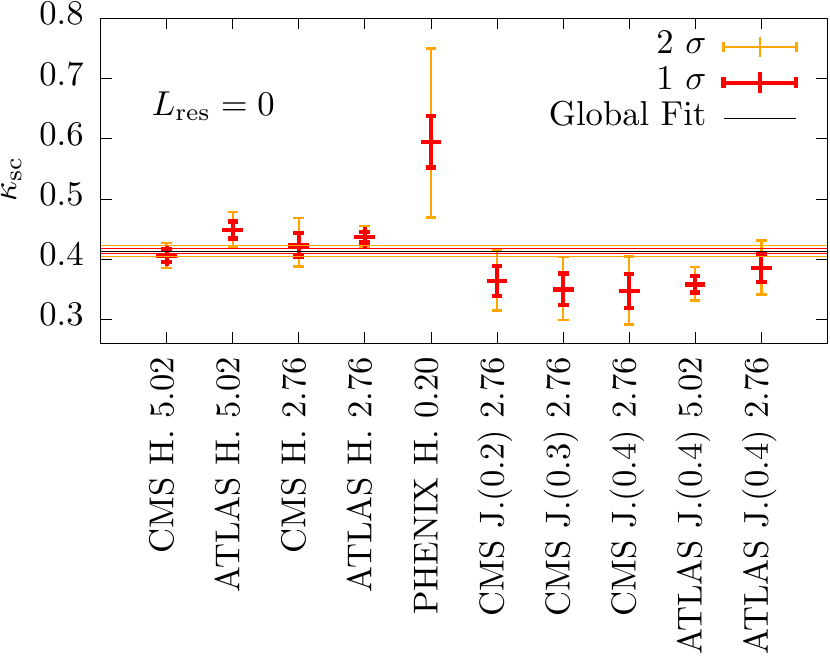}\hfill
	\vspace{-0.1in} 
	\includegraphics[scale=0.85]{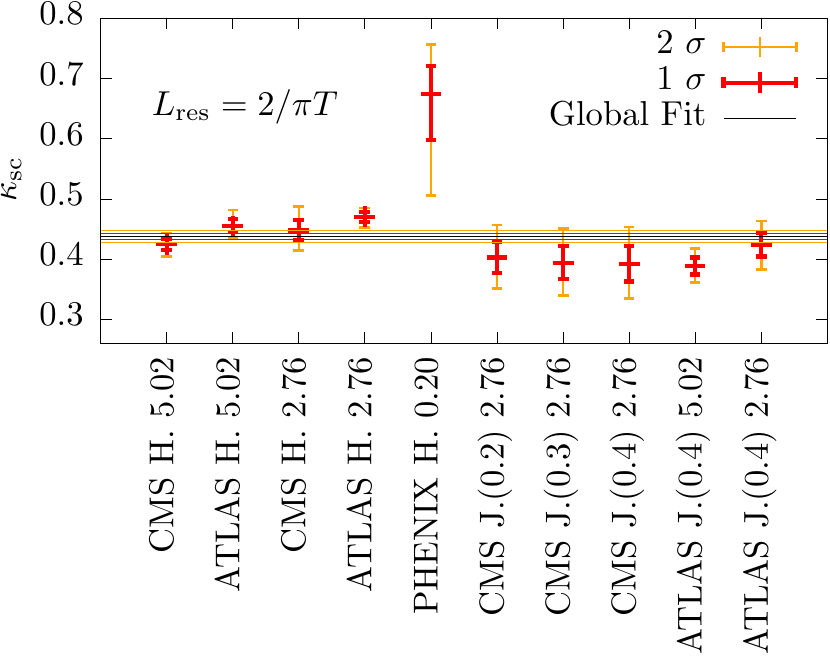}
	\hspace*{\fill}
	
	\caption {Results for the best $\aSC$ after a $\chi^2$ fit, for either each of the ten different sets of data or all LHC data at the same time (global fit), for no resolution effects (left panel) and resolution effects with $\lres=2/\pi T$ (right panel). ``H." stands for charged hadrons (LHC, PbPb collisions) or $\pi^0$ (PHENIX, AuAu collisions) in the $0\textnormal{-}5\%$ centrality bin, while ``J." stands for reconstructed jets, with the anti-$k_T$ radii in parentheses, in the $0\textnormal{-}10\%$ centrality bin.}
	
	\label{fitplots}
\end{figure}

We present in Fig.~\ref{fitplots} the results for the extraction of $\aSC$ without (left panel) and with (right panel) resolution effects with $\rres=2$. We perform the fits in two alternative ways: an extraction of $\aSC$ for each of the ten different sets of data (PbPb collisions with $\sqrt{s}=2.76$ ATeV or $5.02$ ATeV at LHC and AuAu collisions with $\sqrt{s}=200$ AGeV at RHIC), which are shown as ten different values of $\aSC$ in each panel, and a global fit using all data points except the PHENIX ones, shown as a single horizontal line. The best value of $\aSC$ is found by a $\chi^2$ analysis, with different sources of experimental uncertainty (statistical, uncorrelated systematic, correlated systematic, and normalization) accounted for appropriately. The uncertainty bands on the best fit correspond to the the values of $\aSC$ for which $\chi^2=\chi_{\rm min}^2\pm1$ (1 $\sigma$) and $\chi^2=\chi_{\rm min}^2\pm4$ (2 $\sigma$).

As it is clear from Fig.~\ref{fitplots}, the hybrid model can simultaneously describe hadron and jet suppression with a single value of $\aSC$, obtaining $\chi^2_{\rm min}/\rm{d.o.f.}\simeq 1$ for the global fit results. While the preferred value of $\aSC$ is consistent between hadrons and jets in each panel, some statistical tension is reduced once finite resolution effects are included (right panel of Fig.~\ref{fitplots}). Indeed, jet suppression is reduced (having to select higher values of $\aSC$) whenever the effective number of energy loss sources goes down. It is also worth noting that although there is a clear tension between the preferred value of $\aSC$ in the plasma produced in RHIC and LHC collisions, suggesting a stronger coupling at RHIC, this tension is only at the 3 $\sigma$ level.

\section{The role of jet fragmentation functions}

The measured jet fragmentation functions (FFs) \cite{ff_atlas,ff_cms} count the number of hadrons, per jet, carrying an energy fraction $z$ with respect to the total jet energy. Both CMS and ATLAS have shown that there is an excess of soft hadrons, the region around low $z$, in quenched jets compared to those measured in pp collisions. Such low $z$ enhancement has been attributed to different, although potentially co-exisitng, physical mechanisms in the literature, namely the loss of colour coherence \cite{iancu} and the effect of medium response \cite{chanwook}. However, it is not the low $z$ region that is of interest for our current analysis.

\begin{figure}[t]
	\centering 
	\includegraphics[scale=0.83]{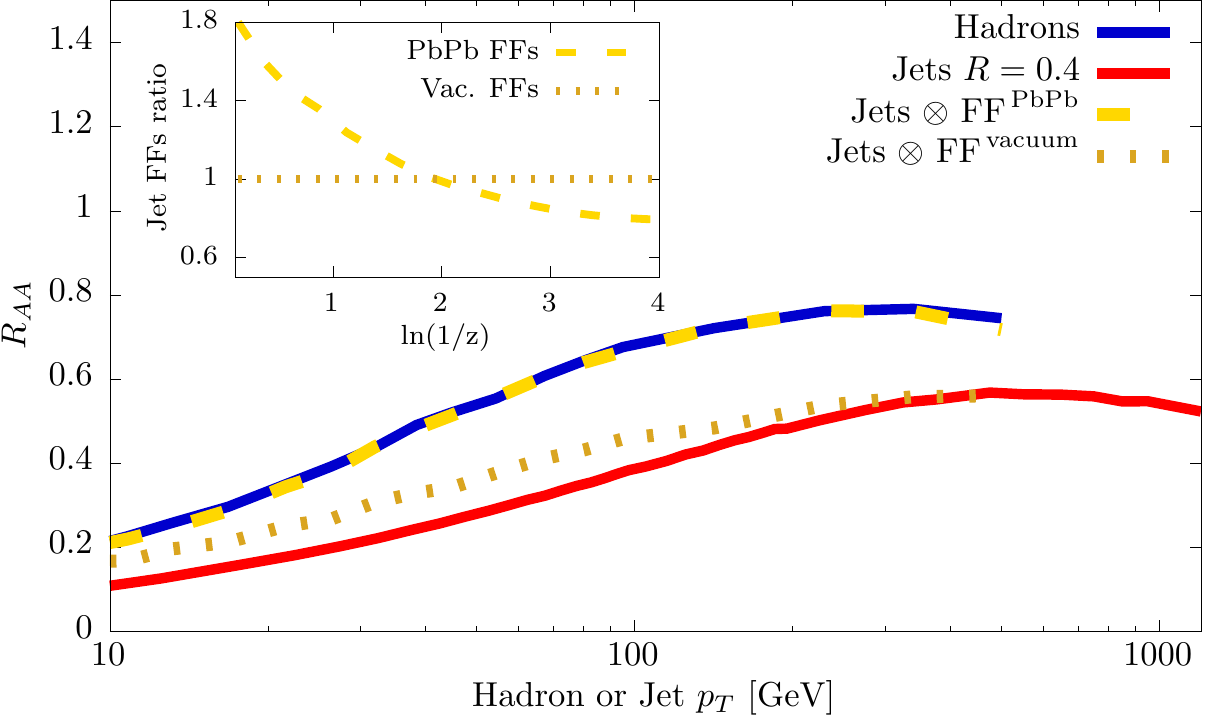}
	
	\caption {Generic hybrid model  calculations of $\jraa$ (red) and $\hraa$ (blue) at LHC, together with two different convolutions of the jet spectra with the FFs: the first, by consistently using the FFs natural to the hybrid model results (dashed yellow), which trivially produce the correct $\hraa$ result, and the second, which wrongly assumes that quenched FFs are the same as vacuum ones (dotted brown).}
	\label{conv}
\end{figure}

The fact that the jet spectrum is steeply falling as a function of $p_T^{\rm jet}$ means that a hadron with any given large $p_T^{\rm hadron}$ is most likely to be a high-$z$ hadron that carries a significant fraction of the momentum of the jet in which it finds itself.  (If it had a smaller $z$, that would mean that the jet of which it is a part had a larger total $p_T^{\rm jet}$, which is less likely.) This means that the hadron spectrum is dominated
by those tracks that contribute to the high-$z$ region of the FFs.  This further implies that high energy hadrons belong, on average, to hard fragmenting jets that are narrow, which lose the least energy as they
propagate through the plasma. Wider jets that started out with some higher $p_T$ and lost more energy are less numerous and do not contribute significantly to the hadron spectrum. To confirm this understanding, we show in Fig. 2 that it is possible to recover the high-$p_T$ hadron spectrum by convolving the jet spectrum with the corresponding FFs.
%
The following are results from the hybrid model exclusively, without any external input. The red curve corresponds to $\jraa$ for $R=0.4$ jets, while the blue one is $\hraa$. Therefore, the $x$ axis corresponds to either jet or hadron $p_T$. By knowing the number of jets per jet $p_T$ bin (the jet spectrum), and using the appropriately binned jet $p_T$ FFs, one can straightforwardly count the number of hadrons per hadron $p_T$ (the hadron spectrum). In this way, by convolving the quenched jet spectrum with the PbPb FFs, and by convolving the pp jet spectrum with the pp FFs, we plot the ratio of the two as the dashed yellow line in Fig.~\ref{conv}. This curve trivially agrees with the blue one, as it has to. In the inset figure at the top left, the reader will find the shape of the ratio between the PbPb and pp FFs used to recover $\hraa$, plotted as a function of $\rm{ln}(1/z)$. By looking at the dashed yellow FFs ratio, one readily observes a pronounced high $z$ enhancement, which we identify as the feature signalling that high $p_T$ hadrons are less suppressed than jets as a whole, leading to the qualitative statement that $\hraa \, > \, \jraa$. If we \emph{wrongly} assumed that quenched FFs are the same as vacuum ones, i.e. if the FFs ratio were one (the brown dotted curve in the top left inset figure), the convolved $\hraa$ (the dotted brown curve in the main figure)  would look much more like $\jraa$ (modulo the expected spectrum shift corresponding to $\langle z \rangle \sim 0.3$).

\section{Conclusions}

The hybrid strong/weak coupling model describes hadron and jet suppression well, simultaneously. Wider jets lose more energy because they contain a greater number of energy loss sources. This fact, together with the steeply falling jet spectrum, biases the final jet distribution towards narrower ones, which contain the hadrons that dominate at high $p_T$. This is reflected in the high $z$ enhancement in the FFs observed both in data \cite{ff_atlas} and in our model, and is tantamount to the observation that $\hraa \, > \, \jraa$ at high $p_T$. 
The number of active components in a jet
effectively depends on the power of the QGP to resolve its internal structure, which means that finite resolution effects play an important role in the physics discussed here. 

This work was supported by  grants SGR-2017-754, FPA2016-76005-C2-1-P and MDM-2014-0367, by Funda\c c\~ao para a Ci\^encia e a Tecnologia (Portugal) contracts CERN/FIS-PAR/0022/2017 and IF/00563/2012, by US NSF grant ACI-1550300 and by US DOE Office of Nuclear Physics contract DE-SC0011090.






\bibliographystyle{elsarticle-num}



\end{document}